\documentclass[10pt,a4paper,oneside,onecolumn,titlepage,final,openany]{article}

\usepackage{amsmath}
\usepackage{multicol}
\usepackage[]{algorithm2e}
\usepackage{mathtools}
\newtagform{brackets}{[}{]}
\usetagform{brackets}
\usepackage{amsfonts}
\usepackage{amssymb,wasysym}
\usepackage{graphicx,titlesec}
\usepackage{abstract}
\usepackage{authblk}
\usepackage{caption}
\usepackage{gensymb}
\usepackage{rotating}
\usepackage{wrapfig}
\usepackage[margin=2.5cm,a4paper]{geometry}
\usepackage[sort]{cite}
\usepackage{setspace}
\usepackage{parskip}
\usepackage{microtype}
\usepackage{tensor}
\usepackage[nomarkers]{endfloat}
\usepackage[capitalise]{cleveref}
\usepackage{textcomp}
\newcommand{\murm}{\hbox{\textmu}}

\usepackage{newfloat}
\usepackage{url}
\usepackage{xspace}

\usepackage{xspace}
\newcommand{\carbon}{$^{13}\text{C}$}
\newcommand{\proton}{$^{1}\text{H}$}

\newcommand{\pyruvate}{[1-\carbon]pyruvate\xspace}

\newcommand{\Tstar}{{T$_2$}$^{*}$\xspace}

\newcommand{\SI}[2]{#1\,#2}

\usepackage{marginnote}

\usepackage[disable]{todonotes}
\usepackage[final]{changes}
\definechangesauthor[color=blue]{ }
\usepackage{changebar}

\DeclareDelayedFloatFlavor{sidewaysfigure}{figure}
\DeclareDelayedFloatFlavor{sidewaystable}{table}
\usepackage{fancyhdr}

\title{A 3D-Hybrid-Shot Spiral Sequence for Hyperpolarized {\carbon} Imaging}
\author[1,2]{Andrew Tyler}
\author[1,2]{Justin Y.~C.~Lau}
\author[1]{Vicky Ball}
\author[1]{Kerstin N.~Timm}
\author[1,2]{Tony Zhou}
\author[1,2]{Damian J.~Tyler*}
\author[1,2,3]{Jack J.~Miller*}
\affil[1]{Department of Physiology, Anatomy \& Genetics, University of Oxford}
\affil[2]{Oxford Centre for Clinical Cardiac Magnetic Resonance Research (OCMR), John Radcliffe Hospital, Hedley Way, Headington, OX3 9DU}
\affil[3]{Department of Physics, Clarendon Laboratory, University of Oxford }

\date{\vspace{1 em}\flushleft A 3D-Hybrid-Shot Spiral Sequence for Hyperpolarized {\carbon} Imaging\\\vspace{1em}Word Count: 4980 (Excluding figure captions and abstract). \\\vspace{1em}Corresponding author:\\Dr Jack J. Miller, DPhil\\ Sherrington Building, Parks Road, Oxford, OX1 3PT\\
	Email: jack.miller@dpag.ox.ac.uk\\
	\vspace{1em}*Joint senior authors\\
	\vspace{2 em}
		This work was supported by funding from the Engineering and Physical Sciences Research Council (EPSRC) and Medical Research Council (MRC) [grant number EP/L016052/1]. All authors would like to acknowledge the support of the British Heart Foundation (Senior Fellowship ref. FS/14/17/30634, and the Oxford BHF Centre for Research Excellence ref. RE/13/1/30181) and the support of the UK NIHR. JYCL would like to acknowledge funding from the NIHR Oxford Biomedical Research Centre and support from the Fulford Junior Research Fellowship at Somerville College. JJM would like to acknowledge financial support by a Novo Nordisk Postdoctoral Fellowship scheme run in conjunction with the University of Oxford, and also by St Hugh's and Wadham college.\\
		\vspace{2 em}
	Submitted to \textit{Magnetic Resonance in Medicine} as a full paper
}

\begin{document}

\maketitle

\begin{abstract}
\doublespacing
\emph{Purpose}: Hyperpolarized imaging experiments have conflicting requirements of high spatial, temporal, and spectral resolution. Spectral-Spatial RF excitation has been shown to form an attractive magnetization-efficient method for hyperpolarized imaging, but the optimum readout strategy is not yet known. \\[2em]
    \emph{Methods}: In this work we propose a novel 3D hybrid-shot spiral sequence which features two constant density regions that permit the retrospective reconstruction of either high spatial or high temporal resolution images \emph{post hoc}, (adaptive spatiotemporal imaging) allowing greater flexibility in acquisition and reconstruction. \\[2em]
    \emph{Results}: We have implemented this sequence, both via simulation and on a pre-clinical scanner, to demonstrate its feasibility, in both a \proton\ phantom and with hyperpolarized \carbon\ pyruvate in vivo. \\[2em]
    \emph{Conclusion}: This sequence forms an attractive method for acquiring hyperpolarized imaging datasets, providing adaptive spatiotemporal imaging to ameliorate the conflict of spatial and temporal resolution, with significant potential for clinical translation.\\[2em]
    \end{abstract}

\doublespacing
\pagestyle{fancy}

\section*{Introduction}
Hyperpolarized \pyruvate is a versatile metabolic probe, which has been used extensively to investigate metabolism and pH in health and disease in vivo\cite{golman2006ar, golman2006cr, gallagher2008magnetic}, and is progressing rapidly towards clinical translation \cite{nelson2013metabolic, cunningham2016hyperpolarized, miloushev2018metabolic, grist2019quantifying, stodkilde2019pilot, Rider2020}. Many conditions where metabolic dysregulation is of interest are spatially localized, therefore it would be desirable to image at sufficiently high spatial resolution to appreciate potential metabolic heterogeneities within the tissue{, for instance in the myocardium post infarction to asses viability when planning intervention}. It is also desirable to perform time resolved imaging to observe temporal metabolic dynamics, which may be significantly altered in pathology, for instance, in ischaemic cardiomyopathies\cite{Lau2013}. Furthermore, {a growing body of work advocates for the use of time-course data for the determination of metabolic parameters, such as k\textsubscript{PL} (pyruvate-lactate), either via the fitting of kinetic models\cite{Geraghty2017,Gordon2019} or calculation of area-under-the-curve ratios\cite{Hill2013}, necessitating high quality time-courses with sufficient temporal resolution to properly characterise the shape of the curve. In addition, }owing to the inherent physiological variability in the time between the injection of hyperpolarized pyruvate and its arrival in the organ of interest, the use of dynamic imaging strategies is preferred, to ensure that the acquisition does not miss the initial bolus of injected substrate \cite{durst2015comparison}. {It is therefore desirable to have both high spatial and temporal resolution in the same aquisition, i}n practice, there is a trade-off between spatial and temporal resolution, which thus far must be predefined \emph{a priori} at the time of acquisition.

The finite, decaying, and non-renewable magnetization generated by hyperpolarization necessitates rapid, magnetization-efficient imaging techniques. One common strategy is to combine the magnetization-efficiency of spectral-spatial RF excitations with rapid spiral readouts \cite{mayer2009application,mayer2010vivo,lau2011spectral,wiesinger2012ideal,wang2017single,miller201813}. The first hyperpolarized {\carbon} images of the human heart were acquired with a single-shot spiral trajectory \cite{cunningham2016hyperpolarized}. Single-shot spiral-out readouts efficiently encode a plane of k-space with minimal dead time by starting at the centre of k-space with the energy-rich low spatial frequencies. However, for a given field of view (FOV), the maximum spatial resolution that can be achieved with a single-shot spiral is limited by the readout duration allowed by the magnetic field inhomogeneity of the sample\cite{lau2019imaging} and the \Tstar of the hyperpolarized nuclei in question.

One approach to overcome the limiting effect of \Tstar is to use a multi-shot spiral readout \cite{meyer1992fast}, which segments the acquisition of k-space into multiple readouts, each temporally shorter and therefore less affected by \Tstar decay than an equivalent single-shot spiral readout. However, as each segment does not typically meet the Nyquist criterion, it is not possible to reconstruct images individually from each interleaf at the prescribed FOV without aliasing. Another strategy to shorten the readout is to design variable density spiral (VDS) trajectories \cite{spielman1995magnetic}, where the sampling density is not constant across k-space. Spiral-out VDS designs with a sampling density that decreases as a function of the distance away from the centre of k-space have been used to reduce artefacts due to motion \cite{liao1997reduction} and aliasing \cite{tsai2000reduced}. However, there is an SNR penalty associated with VDS designs due to noise amplification from the weighting of the varying sampling density in the image reconstruction process.

In this work, we demonstrate the feasibility of adaptive spatiotemporal imaging using a hybrid-shot spiral ({HSS}) trajectory which features a constant single-shot density inner region to satisfy the Nyquist criteria up to a minimum desired spatial resolution, followed by a smooth transition to a constant $N$-shot density sub-Nyquist outer region. By truncating each readout to the inner single-shot region (single-shot HSS), low spatial resolution images can be reconstructed for each TR. By combining non-truncated data which encompass a full set of rotation angles from $N$ acquisitions of the same metabolite, alias-free higher spatial resolution images can be reconstructed (full HSS) at lower temporal resolution. This adaptive spatiotemporal feature of the HSS acquisition allows the trade-off between spatial and temporal resolution to be made \emph{post hoc} at reconstruction time.

\section*{Theory} 
\begin{figure}
    \centering
    \includegraphics[scale = 0.8]{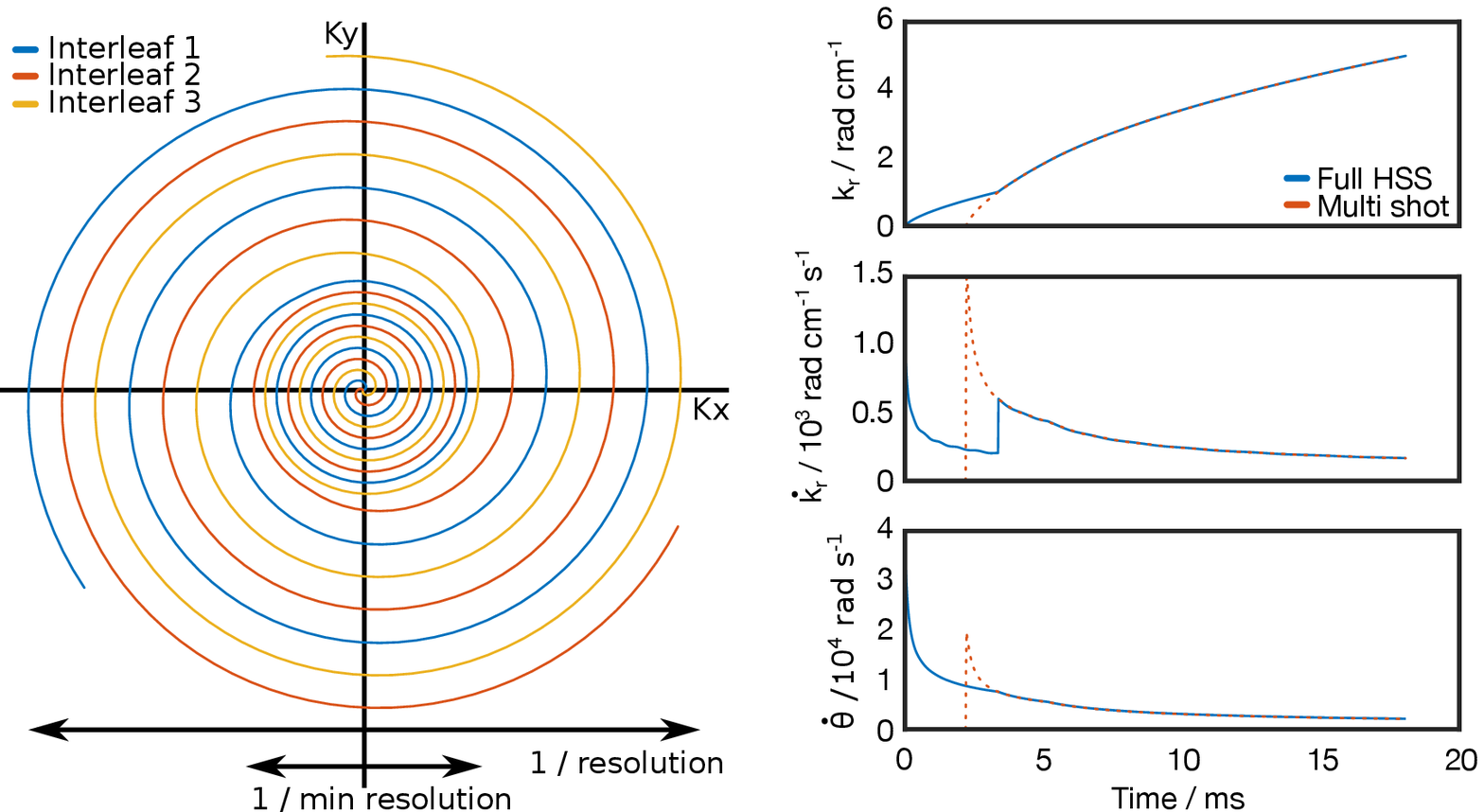}
    \caption{\textbf{Left:} The proposed k-space readout trajectory (experimental parameters as in the digital phantom experiment). In this case three shots are required for a full resolution reconstruction, however an image can be reconstructed every shot at a lower spatial resolution.\newline \textbf{Right:} $k_r$, $\dot k_r$ and $\dot\theta$ values as a function of time for the shown trajectory superimposed with an aligned multi-shot spiral with the same imaging parameters.}
    \label{fig:k_space_traj}
\end{figure}

Numerous strategies for designing spiral trajectories exist in the literature. Given upper limits for gradient amplitude and slew rate, analytical formulae can be derived to design constant density \cite{glover1999simple} and variable density \cite{kim2003simple} spiral trajectories. A hybrid-density trajectory has been proposed for functional MRI with a single-shot inner region up to a fraction of the maximum design k-space radius beyond which the sampling density decreases \cite{chang2011variable}. In this work, the target HSS trajectory{ (shown in Figure \ref{fig:k_space_traj})} consists of two connected regions of constant sampling density with a transition that is smooth but relatively abrupt as opposed to a gradual change in sampling density common with VDS designs using the aforementioned algorithms.

Iterative algorithms compute each point of the trajectory by solving a coupled set of differential equations with the previous point as the initial condition. A widely used example of an iterative VDS design algorithm represents the sampling density as an effective FOV which is a function of k-space radius \cite{lee2003fast,HargreavesTHESIS}. In this algorithm the spiral trajectory is defined as

\begin{equation}
	\mathbf{k} = a\theta e^{i\theta}
	\label{eqn:spiral}
\end{equation}

where $\mathbf{k} = k_x+ik_y$, $\theta$ is the polar angle and $a$ is the rate of increase of k-space radius {($k_r = |\mathbf{k}|$) }with respect to $\theta$, which is defined by

\begin{equation}
	a = \frac{\mathrm{d}k_r}{\mathrm{d}\theta} = \frac{N\Delta k_{max}}{2\pi} = \frac{N}{2\pi \cdot \text{FOV}}
	\label{eqn:a_eqn}
\end{equation}

If FOV is allowed to vary as a function of k-space radius, a variable density spiral can be constructed. Differentiating Eq \ref{eqn:spiral} gives an equation for the gradient vector ($\mathbf{G}$)

\begin{equation}
	\mathbf{G} = \frac{2\pi a}{\gamma}e^{i\theta}(\dot\theta+i\theta\dot\theta)
	\label{eqn:gradient}
\end{equation} 

Which can be further differentiated to give an equation for the slew rate ($\mathbf{S}$)

\begin{equation}
	\mathbf{S} = \frac{2\pi a}{\gamma}e^{i\theta}[(\ddot\theta-\theta\dot\theta^2)+i(2\dot\theta^2 + \theta\ddot\theta)]
	\label{eqn:slew}
\end{equation} 

{By taking the magnitude of Equation \ref{eqn:gradient} and setting $|\mathbf{G}|=G_\text{max}$, the gradient limited value of $\dot k_r$ can be found}

\begin{equation}
	\dot k_\text{max} = \sqrt{\frac{(\gamma\, G_\text{max})^2/4\pi^2}{1 + \left(\frac{k_r}{a}\right)^2}}
\end{equation}

{Similarly, solving Equation \ref{eqn:slew} in the same way, gives an expression for the slew rate limited ($|\mathbf{S}| = S_\text{max}$) value of $\ddot k_r$}

\begin{equation}
	\ddot k_\text{max} = \frac{-B}{2A} + \sqrt{\frac{B^2}{4A^2}-\frac{C}{A}}
\end{equation}

{Where $A$, $B$, and $C$ are the quadratic coefficients used to solve the equation}

\begin{equation}
	A = 1 + \frac{k_r^2}{a^2} 
\end{equation}
\begin{equation}
	B = \frac{2k_r \dot k_r^2}{a^2}
\end{equation}
\begin{equation}
	C = \left(\frac{k_r \cdot \dot k_r^2}{a^2} \right)^2 + \frac{4\dot k_r^4}{a^2}  - \gamma^2 S_{\text{max}}^2/4\pi^2
\end{equation}

{This gives the following equation for $\ddot k_r$, subject to the dual constraints of $G_\text{max}$ and $S_\text{max}$}

\begin{equation}
	\ddot k_r = 
	\begin{cases}
	\ddot k_\text{max} & \dot k_r < \dot k_\text{max}\\
	\left(\dot k_\text{max} - \dot k_r\right) / \Delta t & \dot k_r \geq \dot k_\text{max}
	\end{cases}
\end{equation}

{Where $\Delta t$ is the scanner dwell time. $\ddot \theta$ can then be found with the expression}

\begin{equation}
	\ddot\theta = \frac{1}{a} \cdot \ddot k_r
\end{equation}

{To calculate the trajectory, $\dot\theta$ and $\dot k$ are incremented by $\ddot\theta$ and $\ddot k$ at each timepoint with $\theta$ and $k$ also incremented at each timepoint by the new values of $\dot\theta$ and $\dot k$. All variables are initialised to 0 at the first timepoint and the algorithm is terminated when $k_r$ reaches the value corresponding to the desired resolution.}

The iterative algorithm was modified to design HSS trajectories with a transition from single-shot to $N$-shot sampling density at a k-space radius corresponding to the desired minimum resolution for a high temporal resolution reconstruction. 

{In its modified form, the algorithm begins as for a single shot spiral, then} once the trajectory reache{s} the k-space radius corresponding to the single-shot HSS resolution{, $k_\text{trans}$,} the FOV parameter {is} scaled by a factor of $1/N$, reducing the sampling density to that of a $N$-shot spiral, until the end of the readout. {At the transition between the two density regions, $\dot k_r$ is set equal to $\dot k_\text{trans}${, the value of $\dot{k}_r$ for an $N$-shot spiral at $k_\text{trans}$}. The algorithm is now in a state identical to an $N$-shot spiral at $k_r = k_\text{trans}$, as shown in Figure \ref{fig:k_space_traj}, resulting in a smooth transition between the two densities without violating $G_\text{max}$ or $S_\text{max}$.} {The pseudo-code describing this is\deleted[id= ]{ given below:}}\added[id= ]{ shown in the appendix.}

\color{black}

\deleted[id= ]{The algorithm begins by calculating $\dot k_\text{trans}$ (the first while loop), before reseting and calculating the single shot trajectory up to $k_r = k_\text{trans}$ (the second while loop), at which point $N$ is set to the number of shots in the multi-shot region and $\dot k_r$ is set to $\dot k_\text{trans}$. After which, the algorithm proceeds to calculate the multi-shot portion of the trajectory (the third while loop).}\todo{R2.1}

\section*{Methods} 

\subsection*{Sequence design}

\begin{figure}
    \centering
    \includegraphics[scale = 0.8]{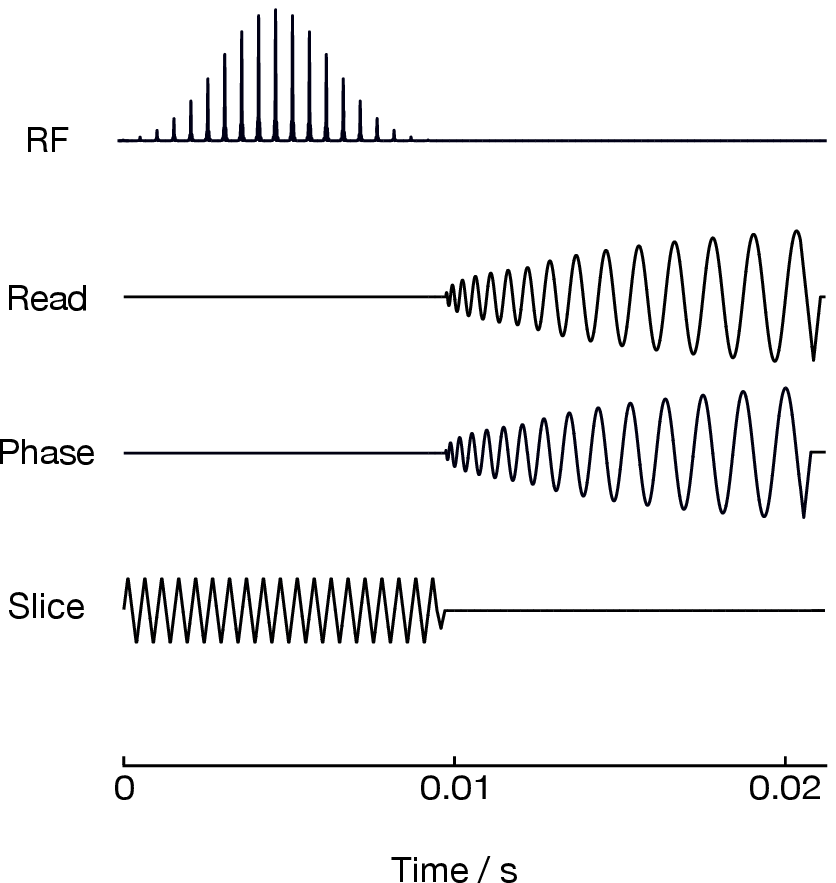}
    \caption{A simplified pulse sequence diagram for a HSS in vivo acquisition showing the spectral-spatial excitation (0--0.01 s), z phase encode (0.01 s, slice) and HSS readout (0.01 s -- end). }
    \label{fig:pulse_sequence}
\end{figure}

The proposed sequence is a 3D hybrid-shot spiral with spectral spatial excitation, centric phase encodes in the $z$-direction and a hybrid spiral readout{ with rewind}. A simplified representation of the sequence is shown in Figure \ref{fig:pulse_sequence}.

{A spectral spatial excitation scheme was used to provide spectrally selective excitation within the volume of interest, while minimising excitation outside of the volume of interest, in order to preserve the pool of hyperpolarised magnetisation circulating in the animal, as our coil could not provide enough spatial localisation by itself.} As described previously \cite{miller2016robust}, {the }spectral spatial excitation pulses were designed under the RF excitation k-space formalism with zipper correction \cite{lau2010rapid}. Sub-pulses under a Gaussian envelope were designed with the Shinnar-Le-Roux algorithm \cite{shinnar1989synthesis}, with a time-bandwidth product of $3$ in the spectral domain and $9$ in the spatial domain, and zero amplitude on negative gradient lobes to produce a fly-back pulse. The scheme was originally designed to be slew-rate ($S_\text{max}$) limited rather than {gradient} ($G_\text{max}$) limited, as we have empirically found at both 7T and 3T that waveforms played at $S_\text{max}$ were reproduced with significantly better fidelity than those at $G_\text{max}${\cite{miller2016robust}}, when measured by the method proposed by Duyn \emph{et al.} \cite{duyn1998simple}.

The pulse was \SI{9}{ms} long, and possessed 37 sub-lobes each of duration \SI{246}{\murm s}, corresponding to an excitation bandwidth of \SI{240}{Hz}, and a stop-band of $\sim$\SI{2}{kHz}, sufficient to avoid every unwanted resonance in the spectrum of hyperpolarized [1-\carbon]pyruvate at 7T, with an excitation bandwidth sufficiently large (\SI{3}{ppm}) to provide relative immunity to transmitting off-resonance and $B_0$ inhomogeneity. The slab thickness was \SI{45.5}{mm}.

In the proposed hybrid-shot spiral k-space encoding scheme, a single-shot and $N$-shot spiral were compounded together and rotated by $360/N^\circ$ so that the central region of k-space was fully sampled every shot and the outer region was fully sampled every $N$ shots. This allowed for a low spatial resolution image to be reconstructed every shot (single-shot HSS), giving maximum temporal resolution, or for $N$ consecutive shots to be combined giving a higher spatial resolution, albeit at the cost of temporal resolution (full HSS).{ The spiral readouts were rotated by $360/N^\circ$, as opposed to the golden angle, in order to maximise the SNR of the reconstructed image when $N$ shots were combined\cite{Winkelmann2007}.} The sequence was implemented using a modified version of the Hargreaves variable density spiral algorithm\cite{lee2003fast} to generate the two spirals which are smoothly transitioned between. 

\subsection*{Image reconstruction}

Image reconstruction was performed with custom software written in the Matlab programming language (Mathworks, Natick, MA, 2018b). It used the Berkeley Advanced Reconstruction Toolbox \cite{BART} implementation of the non-uniform Fourier transform (NUFFT) to reconstruct the data in the x-y plane. The k-space trajectory used in the NUFFT calculation was computed by correcting the prescribed gradients with a previously measured {gradient impulse response function}\cite{Vannesjo2013} and numerically integrating the result. 

{In this work the bulk B\textsubscript{0} shift (which causes a `blurring' artefact in image space or `ringing' in the PSF\cite{Block2005} and is hard to avoid, since the pyruvate centre frequency is not known accurately ahead of time) was corrected by a method similar to that used by Lau et al.\cite{Lau2016}, where the images were reconstructed with a range of different demodulation frequencies and the least blurred image taken, however in this work the selection was done manually, rather than with an automated approach. While this demodulation was necessary from an image quality point of view it also provided an estimate of the actual \textsuperscript{13}C centre frequency, validating that this parameter was set correctly at acquisition.}     

\subsection*{Digital phantom}

To provide initial validation of the sequence, single-shot, multiple-shot and HSS \carbon\ acquisitions were simulated with a synthetic brain phantom\cite{Lejeune2012}. The simulated scanner had $G_\text{max}=\SI{500}{\text{mT/m}}$, $S_\text{max}=\SI{ 536 }{\text{mT/m/ms}}$
and a bandwidth of $\SI{62.5}{\text{kHz}}$. The multiple-shot and HSS trajectories had three interleaves. The trajectories were calculated to achieve an x/y resolution of $\SI{1}{\text{mm}}$ (single-shot HSS $\SI{5}{\text{mm}}$) and field of view of $\SI{60}{\text{mm}}$. After calculation the trajectories were truncated to the length of the multiple-shot spiral ($\SI{15.9}{\text{ms}}$) for comparison with equal readout time, giving resolutions for each method of: single-shot HSS 5 mm, full HSS 1.1 mm, single-shot spiral 1.9 mm and multiple-shot spiral 1 mm. The \Tstar of the simulated phantom was chosen to be $\SI{19}{\text{ms}}$ as a representative value for hyperpolarized pyruvate in vivo and the simulated excitation flip angle, for depletion of the pool of hyperpolarized magnetization, was $3^\circ$. 

Two factors affecting the quality of the reconstructed images were simulated: receive off resonance and noise. Receive off resonance was simulated by multiplying the simulated FID by a phase term corresponding to 0-100 Hz off resonance. Noise was simulated by injecting random Gaussian noise, with a FWHM of 0-10\% the maximum FID intensity, into the simulated FID prior to reconstruction. A structural similarity index (SSIM) was calculated for each reconstruction, using the ground truth phantom as a reference, to assess the reconstructed image quality.   

\subsection*{\textsuperscript{1}H phantom} 

Further validation of the sequence was performed with \proton\ phantom measurements. The resolution phantom consisted of a tube containing Lego pieces filled with a Ferumoxytol solution (elemental iron concentration of \SI{30}{\murm g/mL}) to give a \Tstar of $\SI{12.1}{\text{ms}}$. The scanner was a Varian 7T DDR system with $G_\text{max}=\SI{1000}{\text{mT/m}}$, $S_\text{max}=\SI{ 5000 }{\text{mT/m/ms}}$, coupled with a \SI{72}{mm} dual-tuned proton/carbon birdcage volume coil. A $5^\circ$ sinc pulse with duration \SI{1000}{\murm s} and slab thickness \SI{45.5}{mm} was used for the excitation and the bandwidth was set to \SI{250}{kHz}.    

Single-shot, multiple-shot and HSS 3D acquisitions were acquired for comparison. The multiple-shot spiral and full HSS trajectories each had 3 interleaves. The trajectories were designed with a field of view of $80\times80\times45.5$ mm and resolution of $0.5\times0.5\times1.4$ mm with the single-shot HSS reconstruction giving a resolution of $2.5\times2.5\times1.4$ mm. The trajectories were then truncated to the same readout time (\SI{14.8}{ms}), corresponding to a full HSS resolution of $\SI{0.8}{\text{mm}}$ (multi-shot spiral \SI{0.7}{mm}, single-shot spiral \SI{1.3}{mm}). $32\ z$ phase encode steps were used.  

In addition to the spiral acquisitions, a 3D gradient echo sequence was used to acquire a high resolution ground truth image. The FOV was $80\times80\times45.5$ mm and the resolution $0.4\times0.4\times1.4$ mm. SSIM indices were calculated for each spiral reconstruction over a region of interest (indicated in Figure \ref{fig:proton_phantom_results}) and an intensity profile was plotted through the phantom for each acquisition.

\subsection*{Preclinical imaging}

Preclinical proof-of-concept experiments were performed on a Varian 7T DDR system with $G_\text{max}=\SI{1000}{\text{mT/m}}$, $S_\text{max}=\SI{ 5000 }{\text{mT/m/ms}}$, with an actively detuned transmit/surface receive setup consisting of a \SI{72}{mm} dual-tuned proton/carbon birdcage volume coil with a \SI{40}{ mm} two-channel \carbon~ surface receive array with an integrated preamp (Rapid Biomedical GmbH, Rimpar, Germany).

Approximately \SI{40}{mg} of [1-\carbon]pyruvic acid doped with \SI{15}{mM} OX063 trityl radical and \SI{3}{$\murm$L} of a 1-in-50 dilution of Dotarem (Guerbet Laboratories Ltd) gadolinium chelate was polarized and dissoluted in a prototype polariser as described previously \cite{ardenkjaer2003increase}. Male Wistar rats were anaesthetised with $2\%$ isoflurane in $90\% \text{ O}_2$, $10\% \text{ N$_2$O}$, placed prone onto the imaging coil in a home-built preclinical imaging cradle, and injected with \SI{2}{ml} of \SI{80}{mM} hyperpolarized \pyruvate over approximately \SI{20}{s}.

A thermally polarized \SI{5}{M} {\carbon} urea phantom was included next to the animal to provide a carbon frequency reference. Three dimensional, cardiac-gated volume shimming was performed using the previously mapped behaviour of the magnet's shim set and a spherical harmonic cardiac auto-shimming algorithm as described previously \cite{schneider2009automated}. The \carbon{} centre frequency was set by identifying the carbon resonant frequency of the urea phantom\deleted{and compensating for $B_0$ variation between the urea phantom and the heart by a constant $\Delta B_0$ (\SI{296}{Hz}), on the basis of previously performed spectroscopic experiments.}\added{, and applying a constant offset of \SI{296}{Hz} (\SI{3.95}{ppm}). This consisted of the $B_0$ variation between the urea phantom and the heart (estimated from previously performed $\Delta B_0$ maps), the known chemical shift difference from {\carbon} urea to \pyruvate, and empirical optimisations. }\todo{R2.2}The \carbon{} centre frequency was verified during reconstruction as described in the image reconstruction section.

Once injected with the hyperpolarized \pyruvate solution, {three }rats were imaged using the proposed sequence, consisting of a spectral spatial excitation and {a cardiac gated (one shot per RR interval) }3 interleaf HSS readout scheme. The imaging parameters{, for the first two rats,} consisted of a readout bandwidth of \SI{250}{kHz}, a field of view of $120\times120\times\SI{45.5}{\text{mm}}$, 12 $z$ phase encode steps, a resolution of $2\times2\times\SI{3.8}{\text{mm}}$ and a single-shot HSS resolution of $10\times10\times\SI{3.8}{\text{mm}}${, the nominal minimum TR (time for 1 rotation angle for one metabolite) before gating was \SI{212.4}{ms}}.{ The excitation flip angles were: pyruvate $3^\circ$, bicarbonate $20^\circ$, and lactate $20^\circ$. }One {of these }rat{s} was injected with hyperpolarized \pyruvate a second time in a single session, and imaged with a single-shot spiral with the same FOV and resolution as the full HSS acquisition{ and a nominal TR of \SI{352.0}{ms} before gating}.{ The third rat was imaged with a readout bandwidth of \SI{62.5}{kHz}, a field of view of $60\times60\times\SI{45.5}{\text{mm}}$, 12 $z$ phase encode steps, a resolution of $1\times1\times\SI{3.8}{\text{mm}}$ and a single-shot HSS resolution of $5\times5\times\SI{3.8}{\text{mm}}$}.

After the hyperpolarized imaging, a set of anatomical images were acquired using a gradient echo cine sequence described previously \cite{schneider2003fast}, with the same slice thickness and FOV as the hyperpolarized data.

All experiments were performed in accordance with relevant UK legislation (with personal, project and institutional licences granted under the Animals (Scientific Procedures) Act 1986), and were subject to local ethical review and an independent cost-benefit analysis.

\section*{Results}

\subsection*{Digital phantom}

\begin{figure}
	\centering
	\includegraphics[scale = 0.8]{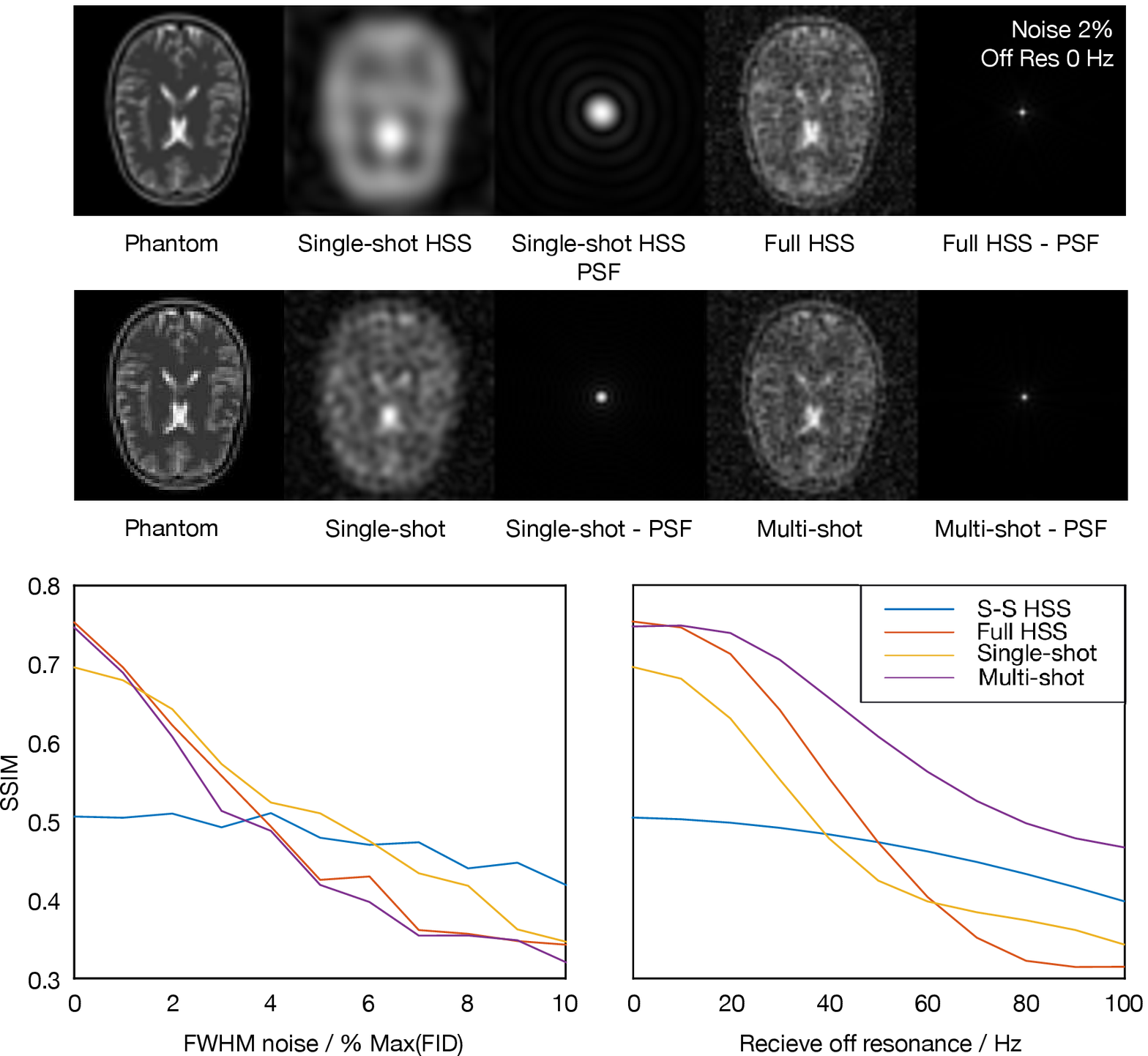}
	\caption{\textbf{Top:} Sample images showing reconstructed HSS, single-shot and multiple-shot acquisitions and a ground truth digital phantom with Gaussian noise (FWHM of noise 2\% maximum value of FID) injected into the FID. \newline\textbf{Bottom:} SSIM of the reconstructed images, with the digital phantom as ground truth, as a function of injected noise and simulated receive off resonance.}
	\label{fig:digital_phantom_results}
\end{figure}

The results of the digital phantom experiment are summarised in Figure \ref{fig:digital_phantom_results}. With no injected noise or simulated off resonance, the full HSS and multiple-shot spiral reconstructions have a similar SSIM metric when using the ground truth as a reference image. The single-shot reconstruction had a lower SSIM than either the multiple-shot or full HSS reconstructions and the single-shot HSS reconstruction had the lowest SSIM. This broadly follows the theoretical resolution achieved by each readout in the acquisition time, although the multiple-shot spiral would have been expected to perform better than full HSS, without the influence of \Tstar, the depletion of the hyperpolarized magnetization pool and oversampling of the centre of k-space for full HSS. 

As noise is added the SSIM for all reconstructions decreases, full HSS, multiple-shot and single-shot, which all have equal readout time, decrease at roughly the same rate. As the injected noise increases the single shot spiral performs better than either the full HSS or multiple-shot spiral. Single-shot HSS, with a shorter readout time than the other reconstructions, decreases less quickly than the other methods so that it has the highest SSIM once the noise FWHM reaches around 6\% the maximum value of the FID. 

The SSIM of all reconstructions decreases with increasing off resonance. The single-shot HSS reconstruction decreases much less quickly than the other readouts, due to its shorter length; however there are differences in the other reconstructions, which each have equal readout time. Of the multiple-shot, single-shot and full HSS reconstructions, the multiple-shot image quality degrades the least quickly with added off resonance. Initially the full HSS reconstruction out performs the single-shot spiral; however at greater than \SI{61.5}{Hz}, this reverses and full HSS gives the lowest SSIM of all the reconstructions.

\subsection*{\textsuperscript{1}H phantom}

\begin{figure}
    \centering
    \includegraphics[scale = 0.8]{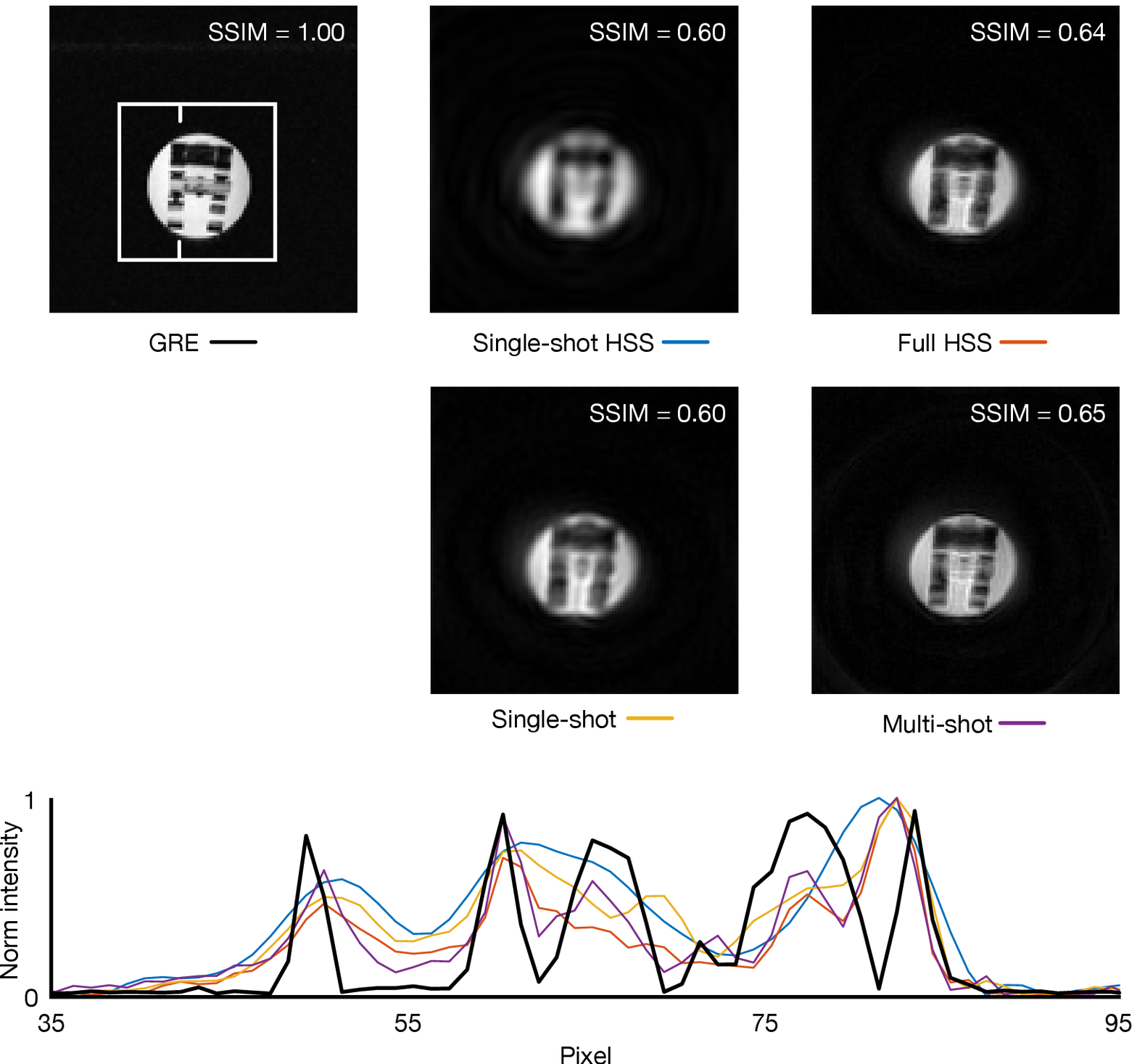}
    \caption{\textbf{Top:} Reconstructed images of a resolution phantom (\Tstar = 12 ms) with structural similarity index, over the area in the white box, with the gradient echo (GRE) acquisition shown as reference, in the top right of each image.\newline\textbf{Bottom:} Normalised line plot through each of the reconstructed images. The position of the line is denoted by the white notches in the gradient echo image.}
    \label{fig:proton_phantom_results}
\end{figure}

Reconstructed images and a line plot through the \proton\ phantom are shown in Figure \ref{fig:proton_phantom_results}. The structural similarity of the reconstructed images to a reference gradient echo image were calculated and broadly followed the expected trend of decreasing SSIM with decreasing resolution. The single-shot HSS reconstruction had a SSIM equal to the single-shot spiral despite achieving a lower resolution potentially due to the shorter readout time reducing the measured noise. 

The line plot through the \proton\ phantom illustrates the trend quantified by the SSIM indices. The multiple-shot and full HSS reconstructions show more definition than the single-shot or single-shot HSS reconstructions which are comparatively blurred. 

Significantly, the HSS acquisition achieved similar image quality to the multiple-shot spiral, when a full reconstruction was performed with the same number of shots, and when only a single-shot was used it produced images of comparable quality to the single-shot spiral while matching its temporal resolution.

\subsection*{Preclinical imaging}

\begin{figure}
    \centering
    \includegraphics[scale=0.8]{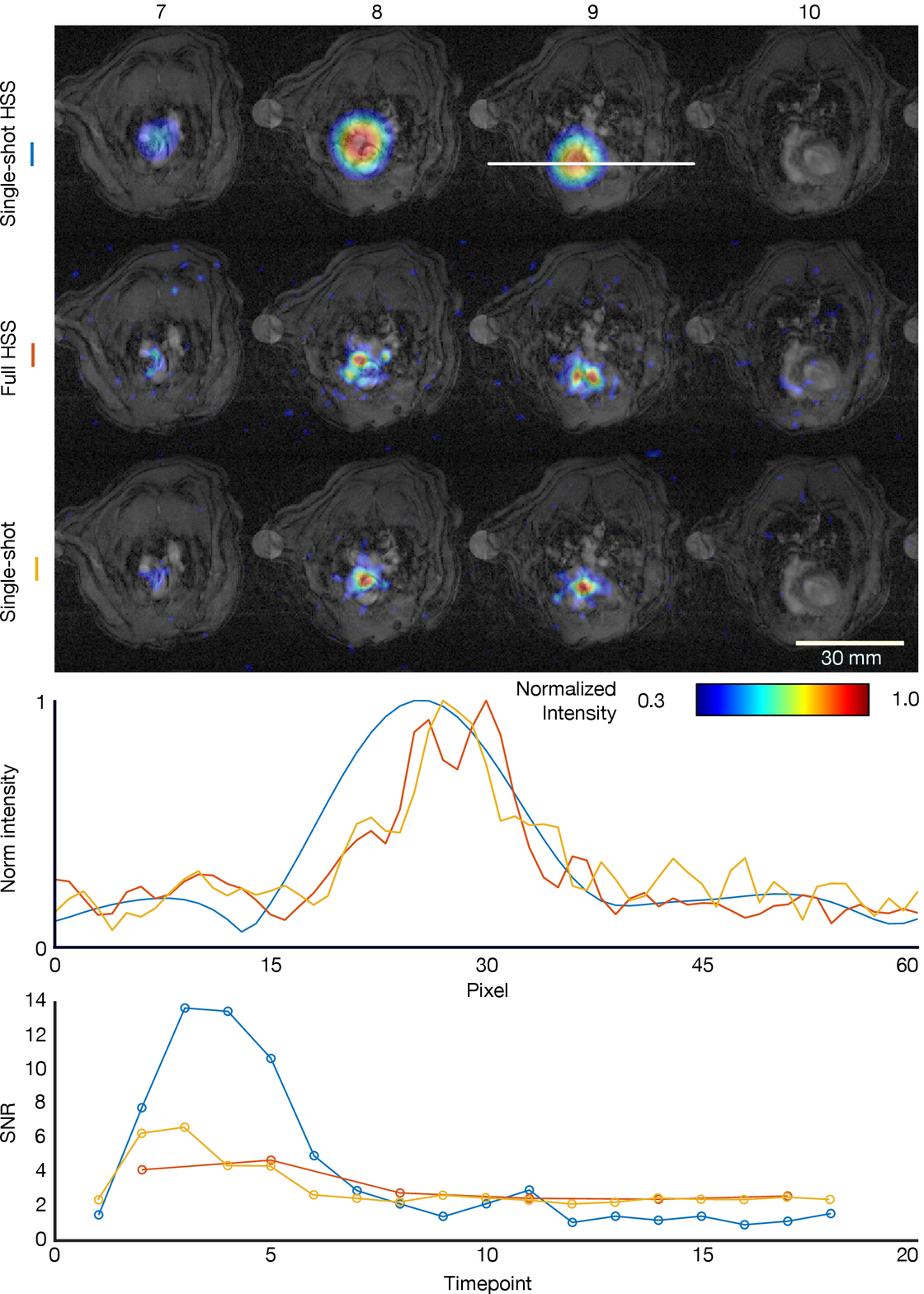}
    \caption{\textbf{Top:} Axial slices 7-10 of HSS and single-shot \carbon\ acquisitions of a rat injected with hyperpolarized \pyruvate overlaid on to matched \proton\ anatomical images, excited at the pyruvate resonant frequency.\newline\textbf{Middle:} Line plots through slice 9 of each reconstruction in the position indicated by the white line in the single-shot HSS image.\newline\textbf{Bottom} SNR of pyruvate in the rat heart (Figure \ref{fig:HYSSvsSS}, slice 9) as a function of acquisition timepoint for single-shot and full reconstructions of a HSS acquisition in addition to a single-shot acquisition.}
    \label{fig:HYSSvsSS}
\end{figure}

\begin{figure}
    \centering
    \includegraphics[scale = 0.8]{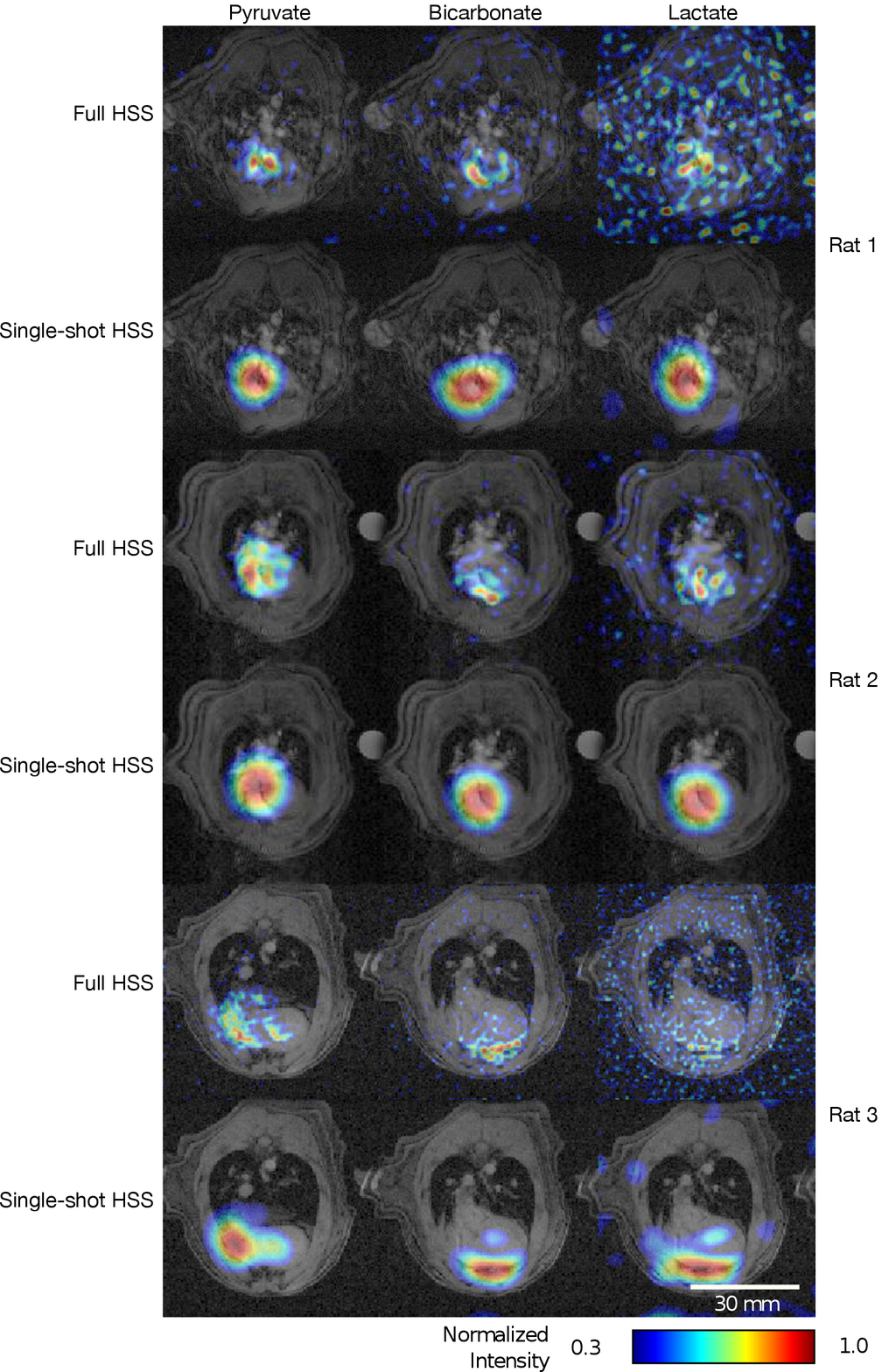}
    \caption{Pyruvate, bicarbonate and lactate HSS images for the rat shown in Figure \ref{fig:HYSSvsSS} (Rat 1) in addition to two additional rats, one (rat 2) imaged with the same imaging parameters and the other (rat 3) with different parameters (bandwidth = 62.5 kHz, FOV = 60 mm, full HSS resolution = 1 mm, single shot HSS resolution = 5 mm).}
    \label{fig:pbl}
\end{figure}

The results of the pre-clinical hyperpolarized experiments are illustrated in Figures \ref{fig:HYSSvsSS} and \ref{fig:pbl}. The line plot in Figure \ref{fig:HYSSvsSS} shows the resolution trade-off made by the HSS sequence. The single-shot HSS reconstruction has a visibly lower {spatial }resolution than the single-shot spiral, however the full HSS reconstruction, separated the left and right ventricles, demonstrating its higher resolution. Figure \ref{fig:HYSSvsSS} {also }shows a time-course of pyruvate SNR for the same hyperpolarized {datasets.} In this example, the single-shot HSS reconstruction captured dynamic information about the system, particularly the bolus arrival and second pass of the pyruvate, while the full HSS reconstruction lacked the temporal resolution to do so. Figure \ref{fig:pbl} shows pyruvate, bicarbonate and lactate metabolic images acquired with the HSS sequence for {three} rats{, two }imaged with the same imaging parameters{ and a third with $4\times$ lower bandwidth, $0.5\times$ FOV and $2\times$ resolution}. In {all three} sets of images the signal is localised to anatomically plausible regions, with the full HSS reconstruction revealing anatomical detail, which was not resolved by the single-shot HSS reconstruction.  
 
\section*{Discussion}

In this work we demonstrated the feasibility of the HSS pulse sequence for use in hyperpolarized \pyruvate\ cardiac imaging. We also explored the trade-offs which inform the choice of spiral readout when designing hyperpolarized studies. Due to the inherent difficulty in measuring image quality for hyperpolarized reconstructions, as there is no ground truth, we performed validation work in both digital and \proton\ phantoms in addition to demonstrating feasibility in a pre-clinical rat model.

The main features of the sequence, and the trade-offs it makes, are illustrated clearly by the digital phantom simulation data presented here. In the ideal case, the full HSS reconstruction achieved similar image quality to a multiple-shot readout with the same readout duration, while retaining the ability to reconstruct a low resolution image from a single shot. The main trade-off of the sequence, the single-shot HSS resolution, is clearly demonstrated by its lower SSIM than the single-shot spiral simulation.  

Furthermore, the simulated full HSS sequence had broadly similar robustness to noise compared to both single and multiple-shot readouts. However as injected noise increased the single shot spiral outperformed the full HSS and multiple shot readouts, potentially due to it incorporating less noise from a shorter total (across all interleaves) readout duration.

Of the equal readout length simulations, (i.e. not single-shot HSS) the multiple-shot simulation had the greatest resilience to off resonance effects, with the full HSS simulation initially performing better than the single-shot simulation before degrading more quickly. Since the readout times were the same, it cannot account for this difference, however we believe that it could potentially be in part due to the re-gridding process in the NUFFT and the relative amount of time spent during the readout at high spatial frequencies away from the centre of k-space. 

The \proton\ phantom provided quantitative validation of the simulations in a controlled experimental system. While closely mirroring the digital phantom simulations, the similarity in SSIM metric between the single-shot spiral and single-shot HSS acquisition was surprising, however from the line plot through the phantom it is clear that the single-shot spiral did achieve a higher spatial resolution despite a poorer similarity to the reference image, potentially due to greater noise. 

The pre-clinical results reported here are qualitatively consistent with the digital and \proton\ phantom validation data, and further illustrate the characteristics of the sequence. The improved resolution of a full HSS acquisition over a single-shot spiral with equal prescribed resolution is demonstrated by the line plot through the heart shown in Figure \ref{fig:HYSSvsSS}, where the pyruvate signal is clearly localised to the two ventricles, unlike for the single-shot spiral. In addition the time course data shown in Figure \ref{fig:HYSSvsSS} demonstrates that the single-shot HSS reconstruction is able to capture appreciably more of the dynamics of the system{, neccesary for kinetic modelling,} than the full HSS reconstruction, albeit with a lower spatial resolution than a single-shot spiral with the same temporal resolution. {It is also worth noting that for a single-shot HSS reconstruction, or single-shot spiral each timepoint is reconstructed from consecutive readouts (i.e. for 12 z phase-encodes and a heart rate of 400 bpm, over \SI{1.8}{s}), unlike for a multi-shot spiral, or full HSS, where each timepoint is a moving average of $N$ shots which may be temporally separated by readouts for other metabolites, making data analysis for kinetic modelling more challenging.}

Figure \ref{fig:pbl} shows pyruvate, bicarbonate and lactate images reconstructed from the same data shown in Figure \ref{fig:HYSSvsSS}{,} as well as images acquired from {two further rats}.{ Image quality across the datasets is comparable to prior work imaging the rat heart at 7T\cite{miller2016robust}. Furthermore, the alternative set of imaging parameters, used for rat 3, resulted in noticeably improved image quality, with better delineation of the myocardium. However the drop-off in signal towards the rear of the heart, particularly noticeable in the bicarbonate images, due to the sensitivity profile of the \textsuperscript{13}C surface receive coil, demonstrates the challenge of hyperpolarised imaging of the rat heart at 7T}. These images demonstrate the reproducibility of the sequence, as well as the feasibility of combining the HSS readout with a spectral-spatial excitation, which allows spatial details, such as the bicarbonate signal from the myocardium, to be resolved while maintaining temporal resolution and spectral selectivity.

Overall, the HSS pulse sequence provides an attractive compromise between spatial and temporal resolution for hyperpolarized \carbon\ MRI, by making a spatial resolution trade-off compared to a traditional multiple-shot acquisition for the ability to reconstruct as a single or multiple-shot spiral from a single acquisition. As such, it forms an attractive candidate for use in clinical imaging, where the benefits of the increased resolution it can provide, compared to a single shot spiral, would be appreciated, particularly when imaging conditions with spatially localised metabolic dysregulation such as myocardial infarction.   

\section*{Conclusion}

Hyperpolarized \carbon\ imaging of the human heart has been demonstrated with a single-shot spiral readout, however the maximum resolution, for a given FOV, this approach can achieve is fundamentally limited by \Tstar decay. This limitation may negatively impact the prospects of developing diagnostic methods for conditions such as ischaemia of the heart, which have shown promise in a pre-clinical setting, that rely on resolving tissue metabolism heterogeneity. The HSS sequence aims to address this limitation of spiral imaging by making a spatial resolution trade-off compared to a traditional multiple-shot acquisition for the ability to reconstruct each shot individually. This allows for multiple-shot images to be reconstructed at a resolution greater than is achievable with a single-shot acquisition, while offering the option to reconstruct images at the temporal resolution of a single-shot spiral, with the trade-off of reduced spatial resolution compared to a standard single-shot sequence.

The experiments presented in this work demonstrate the capabilities of the sequence, as well as its relative trade-offs, and suggest the benefits it may have going forward into clinical translation for hyperpolarized \carbon\ imaging. In our future work we initially plan to further develop the sequence by implementing a $B_0$ inhomogeneity correction to reduce the impact, identified by simulations, of off resonance effects in vivo before implementing the sequence on a clinical system for use in human hyperpolarized \carbon\ cardiac studies.

\section*{Appendix -- HSS readout algorithm}

\added[id= ]{Pseudo-code, describing the algorithm used to generate the spiral trajectories in this work, is shown below. The algorithm was modified from an existing VDS design algorithm\cite{lee2003fast} and implemented in the C programming language.}\todo{R2.1}

\begin{algorithm}[H]
$N \leftarrow N$\;
$\ddot k_r$, $\ddot\theta$, $\dot k_r$, $\dot\theta$, $k_r$, $\theta \leftarrow 0$\;
\While{$k_r < k_\text{trans}$}{
		calculate next $\ddot k_r$, $\ddot\theta$, $\dot k_r$, $\dot\theta$, $k_r$, $\theta$\;
		}
$\dot k_\text{trans} \leftarrow \dot k_r$\;

$\ddot k_r$, $\ddot\theta$, $\dot k_r$, $\dot\theta$, $k_r$, $\theta \leftarrow 0$\;
$N \leftarrow 1$\;
\While{$k_r < k_\text{trans}$}{
		calculate next $\ddot k_r$, $\ddot\theta$, $\dot k_r$, $\dot\theta$, $k_r$, $\theta$\;
		output $\ddot k_r$, $\ddot\theta$, $\dot k_r$, $\dot\theta$, $k_r$, $\theta$\;

		}
$N \leftarrow N$\;
$\dot k_{r} \leftarrow \dot k_\text{trans}$\;   
\While{$k_r < k_\text{max}$}{
		calculate next $\ddot k_r$, $\ddot\theta$, $\dot k_r$, $\dot\theta$, $k_r$, $\theta$\;
		output $\ddot k_r$, $\ddot\theta$, $\dot k_r$, $\dot\theta$, $k_r$, $\theta$\;

		}
	
\end{algorithm}

\added[id= ]{The algorithm begins by calculating $\dot k_\text{trans}$ (the first while loop), before reseting and calculating the single shot trajectory up to $k_r = k_\text{trans}$ (the second while loop), at which point $N$ is set to the number of shots in the multi-shot region and $\dot k_r$ is set to $\dot k_\text{trans}$. After which, the algorithm proceeds to calculate the multi-shot portion of the trajectory (the third while loop).}




\end{document}